\documentstyle[12pt,graphicx,pstricks,braket,amsmath,amssymb,enumitem,
                   float, epstopdf,pdfpages,cite,tabu,hyperref,subfigure, float,tikz]{article}
                    \usetikzlibrary{arrows,matrix,positioning}
\numberwithin{equation}{section}

\newcommand{\hi}[1]{}

\begin{document}

\def\AEF{A.E. Faraggi}

\def\JHEP#1#2#3{{JHEP} {\textbf #1}, (#2) #3}
\def\vol#1#2#3{{\bf {#1}} ({#2}) {#3}}
\def\NPB#1#2#3{{\it Nucl.\ Phys.}\/ {\bf B#1} (#2) #3}
\def\PLB#1#2#3{{\it Phys.\ Lett.}\/ {\bf B#1} (#2) #3}
\def\PRD#1#2#3{{\it Phys.\ Rev.}\/ {\bf D#1} (#2) #3}
\def\PRL#1#2#3{{\it Phys.\ Rev.\ Lett.}\/ {\bf #1} (#2) #3}
\def\PRT#1#2#3{{\it Phys.\ Rep.}\/ {\bf#1} (#2) #3}
\def\MODA#1#2#3{{\it Mod.\ Phys.\ Lett.}\/ {\bf A#1} (#2) #3}
\def\RMP#1#2#3{{\it Rev.\ Mod.\ Phys.}\/ {\bf #1} (#2) #3}
\def\IJMP#1#2#3{{\it Int.\ J.\ Mod.\ Phys.}\/ {\bf A#1} (#2) #3}
\def\nuvc#1#2#3{{\it Nuovo Cimento}\/ {\bf #1A} (#2) #3}
\def\RPP#1#2#3{{\it Rept.\ Prog.\ Phys.}\/ {\bf #1} (#2) #3}
\def\APJ#1#2#3{{\it Astrophys.\ J.}\/ {\bf #1} (#2) #3}
\def\APP#1#2#3{{\it Astropart.\ Phys.}\/ {\bf #1} (#2) #3}
\def\EJP#1#2#3{{\it Eur.\ Phys.\ Jour.}\/ {\bf C#1} (#2) #3}
\def\etal{{\it et al\/}}
\def\notE6{{$SO(10)\times U(1)_{\zeta}\not\subset E_6$}}
\def\E6{{$SO(10)\times U(1)_{\zeta}\subset E_6$}}
\def\highgg{{$SU(3)_C\times SU(2)_L \times SU(2)_R \times U(1)_C \times U(1)_{\zeta}$}}
\def\highSO10{{$SU(3)_C\times SU(2)_L \times SU(2)_R \times U(1)_C$}}
\def\lowgg{{$SU(3)_C\times SU(2)_L \times U(1)_Y \times U(1)_{Z^\prime}$}}
\def\SMgg{{$SU(3)_C\times SU(2)_L \times U(1)_Y$}}
\def\Uzprime{{$U(1)_{Z^\prime}$}}
\def\Uzeta{{$U(1)_{\zeta}$}}

\newcommand{\cc}[2]{c{#1\atopwithdelims[]#2}}
\newcommand{\bev}{\begin{verbatim}}
\newcommand{\beq}{\begin{equation}}
\newcommand{\tr}{Tr}
\newcommand{\ba}{\begin{eqnarray}}
\newcommand{\ea}{\end{eqnarray}}

\newcommand{\beqa}{\begin{eqnarray}}
\newcommand{\beqn}{\begin{eqnarray}}
\newcommand{\eeqn}{\end{eqnarray}}
\newcommand{\eeqa}{\end{eqnarray}}
\newcommand{\eeq}{\end{equation}}
\newcommand{\beqt}{\begin{equation*}}
\newcommand{\eeqt}{\end{equation*}}
\newcommand{\Eev}{\end{verbatim}}
\newcommand{\bec}{\begin{center}}
\newcommand{\eec}{\end{center}}
\newcommand{\bes}{\begin{split}}
\newcommand{\ees}{\end{split}}
\def\ie{{\it i.e.~}}
\def\eg{{\it e.g.~}}
\def\half{{\textstyle{1\over 2}}}
\def\nicefrac#1#2{\hbox{${#1\over #2}$}}
\def\third{{\textstyle {1\over3}}}
\def\quarter{{\textstyle {1\over4}}}
\def\m{{\tt -}}
\def\mass{M_{l^+ l^-}}
\def\p{{\tt +}}

\def\slash#1{#1\hskip-6pt/\hskip6pt}
\def\slk{\slash{k}}
\def\GeV{\,{\rm GeV}}
\def\TeV{\,{\rm TeV}}
\def\y{\,{\rm y}}

\def\l{\langle}
\def\r{\rangle}
\def\LRS{LRS  }

\begin{titlepage}
\samepage{
\setcounter{page}{1}
\rightline{LTH--1095}
\vspace{1.5cm}

\begin{center}
 {\Large \bf F-Theory $E_7$, Heterotic String-Derived Vacua And Flipped $SO(10)$ In Hor\v{a}va-Witten Theory}
\end{center}

\begin{center}

{\large
Johar M. Ashfaque$^\spadesuit$\footnote{email address: jauhar@liv.ac.uk}
}\\
\vspace{1cm}
$^\spadesuit${\it  Dept.\ of Mathematical Sciences,
             University of Liverpool,
         Liverpool L69 7ZL, UK\\}
\end{center}

\begin{abstract}
In this paper, $E_6$ and especially $E_7$ GUT are considered in the F-theory setting in view of the free fermionic construction of the $4D$ heterotic string. In particular, the NAHE-Based LRS model of \cite{Cleaver:2000ds, Cleaver:2002ps} is revisited as an illustration where the starting point was taken to be the $N=4$, $E_7 \times E_7 \times SO(16)$  which through the use of boundary condition basis vectors is reduced to obtain the flipped $SO(10)$ GUT symmetry. 

We also seek to extend the results of \cite{Faraggi:2002ah} in the case of the flipped $SU(5)$ to home in on the flipped $SO(10)$ vacua in the Hor\v{a}va-Witten theory where the $E_8$ gauge group on the observable sector decomposes as $E_{8}\supset E_{6}\times SU(3)$ with $E_{6}$ being the gauge group of the effective field theory. We find for the $E_{6}$ GUT symmetry, solutions of type A and solutions of type B where the Hirzebruch surfaces are considered for the base contrary to \cite{Faraggi:2002ah} where flipped $SU(5)$ vacua were studied and only solutions of type B were found. Moreover, no solutions are found in the case of the base being the del Pezzo surfaces. Furthermore, this goes hand in hand with the heterotic, low-energy string-derived effective model discussed in \cite{Ashfaque:2016psv,Ashfaque:2016ydg}.
\end{abstract}
\smallskip}
\end{titlepage}

\section{Introduction}
To break GUT symmetries in F-theory \cite{Vafa:1996xn, Morrison:1996na, Morrison:1996pp} to construct viable models, one can use Wilson lines \cite{Beasley:2008k,Chung:2009ib} or introduce a $U(1)$ flux corresponding to a fractional line bundle \cite{Beasley:2008kw, Marsano:2009ym, Marsano:2009gv, Marsano:2009wr, Chen:2010tp, Chen:2010ts}. 

The basic ingredient for model building in F-theory is a space-time filling seven-brane which wraps a four-dimensional internal subspace of the six internal directions of the compactification where each lower-dimensional subspace provides an important model building element as can be seen from Table \ref{mb}. An Abelian or a non-Abelian gauge flux of the rank greater than two can then be turned on in the bulk to break the gauge group \cite{Beasley:2008kw}. 
\begin{table}[H]\label{mb}
\begin{center}
	\begin{tabular}{|l|l|l|l|}
		\hline
		Dimension & Ingredient & Complex&Enhancements \\
		&&Codimension&\\
		\hline
		$8D$& Gauge Theory& $1$&-\\
		\hline
		$6D$& Matter &$2$&Rank $1$\\
		\hline
		$4D$& Yukawa Couplings&$3$&Rank $2$\\
		\hline
	\end{tabular}
\end{center}
\caption{The key elements of model building in F-theory.}
\end{table}
 
 The aim of this paper is to present a study of the flipped $SO(10)$ model embedded completely in the $E_{6}$ and $E_{7}$ GUT but with a different accommodation of the SM representations in the ${\bf{27}}$ of $E_{6}$ in string-derived, heterotic low-energy effective models constructed in the free fermionic formulation.
  
\subsection{F-Theory $E_6$}
$$E_{8}\supset E_{6}\times SU(3)_{\perp}$$
with 
$${\bf{248}} \rightarrow ({\bf{78}},{\bf{1}})+({\bf{1}},{\bf{8}})+({\bf{27}},{\bf{3}})+(\overline{{\bf{27}}},\overline{{\bf{3}}})$$
where the inhomogeneous Tate form for $E_6$ is given by
$$x^{3}-y^{2}+b_{1}xyz+b_{2}x^{2}z^{2}+b_{3}yz^{2}+b_{4}xz^{3}+b_{6}z^{5}=0.$$
Here, the $SU(3)_{\perp}$ factor is considered as the group `perpendicular' to the $E_{6}$ GUT divisor.
In what follows assume semi-local approach where the $E_{6}$ representations transform non-trivially under the $SU(3)_{\perp}$. In the spectral cover approach the $E_{6}$ representations are distinguished by the weights $t_{1,2,3}$ of the $SU(3)_{\perp}$ Cartan subalgebra subject to the traceless condition 
$$\sum_{i=1}^{3}t_{i}=0$$
while the $SU(3)_{\perp}$ adjoint decomposes into singlets. All the various $E_6$ breaking patterns 
\begin{eqnarray}
(1a)\,\,E_{6} &\rightarrow& SO(10)\times U(1) \rightarrow SU(5)\times U(1)^{2}\\
(1b)\,\,E_{6} &\rightarrow& SO(10)\times U(1)\rightarrow SU(4) \times SU(2) \times SU(2) \times U(1)\\
(2a)\,\,E_{6} &\rightarrow& SU(6) \times SU(2)\rightarrow SU(5) \times SU(2)\times U(1)\\
(2b)\,\,E_{6} &\rightarrow& SU(6) \times SU(2)\rightarrow SU(4) \times SU(2) \times SU(2)\times U(1)\\
(2c)\,\,E_{6} &\rightarrow& SU(6) \times SU(2)\rightarrow SU(3)\times SU(3) \times SU(2) \times U(1)	\\
(3)\,\,E_{6} &\rightarrow& SU(3) \times SU(3) \times SU(3)
\label{e6break}
\end{eqnarray}
reduce to one of the two extended MSSM models of rank $6$
\begin{eqnarray*}
	E_{6}&\rightarrow& SU(3) \times SU(2) \times [U(1)^{3}]\\
	E_{6}&\rightarrow &SU(3) \times SU(2)\times [SU(2)\times U(1)^{2}]\\
\end{eqnarray*}
which are equivalent up to linear transformations.\footnote{see \cite{Chen:2010tg, and:2015uya} for example.}

\subsection{A String-Derived Low-Energy Effective Model}\label{E6Embedding}
The string-derived model presented in \cite{frzprime, Ashfaque:2016psv, Ashfaque:2016ydg, Faraggi:2016xnm, Ashfaque:2016jha, Athanasopoulos:2014bba} was constructed in the free fermionic formulation \cite{fff} of the four-dimensional heterotic string. The details along with the the massless spectrum  and the superpotential can be found in \cite{frzprime} and therefore omitted here. It was shown that the space-time vector bosons are obtained solely from the untwisted sector and generate the observable and hidden gauge symmetries:
\beqn
{\rm observable} ~: &~~~~~~~~SO(6)\times SO(4) \times 
U(1)_1 \times U(1)_2\times U(1)_3 \nonumber\\
{\rm hidden}     ~: &SO(4)^2\times SO(8)~.~~~~~~~~~~~~~~~~~~~~~~~\nonumber
\eeqn

Under the decomposition $E_{6}\rightarrow SO(10)\times U(1)_{\zeta}$, following from Table \ref{table27rot}, the fundamental representation of $E_{6}$ decomposes as follows
\beqn
{\bf{27}} & \rightarrow & {\bf{16}}_{+{1/2}} + {\bf{10}}_{-1} + {\bf{1}}_{+2}.\nonumber
\eeqn

\begin{table}[H]
	\noindent 
	{\small
		\begin{center}
			{\tabulinesep=1.2mm
				\begin{tabu}{|l|cc|c|c|c|}
					\hline
					Field &$\hphantom{\times}SU(3)_C$&$\times SU(2)_L $
					&${U(1)}_{Y}$&${U(1)}_{Z^\prime}$  \\
					\hline
					$Q_L^i$&    $3$       &  $2$ &  $+\frac{1}{6}$   & $-\frac{2}{3}$   ~~  \\
					$u_L^i$&    ${\bar3}$ &  $1$ &  $-\frac{2}{3}$   & $-\frac{2}{3}$   ~~  \\
					$d_L^i$&    ${\bar3}$ &  $1$ &  $+\frac{1}{3}$   & $-\frac{4}{3}$  ~~  \\
					$e_L^i$&    $1$       &  $1$ &  $+1          $   & $-\frac{2}{3}$  ~~  \\
					$L_L^i$&    $1$       &  $2$ &  $-\frac{1}{2}$   & $-\frac{4}{3}$  ~~  \\
					%
					\hline
					$D^i$       & $3$     & $1$ & $-\frac{1}{3}$     & $+\frac{4}{3}$  ~~    \\
					${\bar D}^i$& ${\bar3}$ & $1$ &  $+\frac{1}{3}$  &   ~~$~2$  ~~    \\
					$H^i$       & $1$       & $2$ &  $-\frac{1}{2}$   &  ~~$~2$ ~~    \\
					${\bar H}^i$& $1$       & $2$ &  $+\frac{1}{2}$   &   $+\frac{4}{3}$   ~~  \\
					\hline
					$S^i$       & $1$       & $1$ &  ~~$0$  &  $-\frac{10}{3}$       ~~   \\
					\hline
					$h$         & $1$       & $2$ &  $-\frac{1}{2}$  &  $-\frac{4}{3}$  ~~    \\
					${\bar h}$  & $1$       & $2$ &  $+\frac{1}{2}$  &  $+\frac{4}{3}$  ~~    \\
					\hline
					$\phi$       & $1$       & $1$ &  ~~$0$         & $-\frac{5}{3}$     ~~   \\
					$\bar\phi$       & $1$       & $1$ &  ~~$0$     & $+\frac{5}{3}$     ~~   \\
					\hline
					%
					$\zeta^i$       & $1$       & $1$ &  ~~$0$  &  ~~$0$       ~~   \\
					\hline
				\end{tabu}}
			\end{center}
		}
		\caption{\label{table27rot}
			\it
			The massless spectrum where the charges are displayed in the 
			normalisation used in free fermionic 
			heterotic string-derived models.}
	\end{table}

The $E_{6}$ GUT symmetry can be broken following \cite{Hewett:1988xc, Callaghan:2012rv,  Chen:2010tg, and:2015uya} as
\begin{eqnarray*}
	E_{6}&\rightarrow& SO(10)\times U(1)_{\zeta}\\&\rightarrow& [SU(5)\times U(1)_{\zeta'}]\times U(1)_{\zeta}\\ &\rightarrow& [SU(3)\times SU(2)\times U(1)_{\zeta''}]\times U(1)_{\zeta'} \times U(1)_{\zeta}
\end{eqnarray*}
and the SM representations are accommodated in the ${\bf{27}}$ of $E_{6}$ as 
$${\bf{27}} = \begin{cases} {\bf{16}}_{+\frac{1}{2}} & \mathcal{F}_{L}+\mathcal{F}_{R} = ($$Q$$,\,$$u^{c}$$, d^{c}, L, e^{c}, N)\\& \rightarrow \binom{Q\,\,\,\,u^{c}}{e^{c}}+\binom{d^{c}}{L}+N\\
{\bf{10}}_{-1}& \mathcal{D} + \mathcal{H} \\ \,\,\,{\bf{1}}_{+2} & {\mathcal{S}\rightarrow }\,\,S\end{cases}.$$ 

\subsubsection{The $E_7$ Enhancement: A Word}
The enhancement $E_{6}\rightarrow E_{7}$ where the extra matter
transforms in the ${\bf{27}}$ representation of $E_{6}$ as can be determined from the
branching rule 
$${\bf{133}}\rightarrow {\bf{78}}_{{0}} \oplus {\bf{1}}_{0} \oplus {\bf{27}}_{+1} \oplus \overline{{\bf{27}}}_{-1}$$
where the subscripts denote the $U(1)$ charges under the decomposition
$$E_7 \rightarrow E_{6} \times  U(1).$$

\section{F-Theory $E_{7}$}
$$E_{8}\supset E_{7}\times SU(2)_{\perp}$$
with 
$${\bf{248}} \rightarrow ({\bf{133}},{\bf{1}})\oplus({\bf{1}},{\bf{3}})\oplus({\bf{56}},{\bf{2}})$$
where the inhomogeneous Tate form for $E_7$ is given by
$$x^{3}-y^{2}+b_{1}xyz+b_{2}x^{2}z^{2}+b_{3}yz^{3}+b_{4}xz^{3}+b_{6}z^{5}=0.$$
\subsection{{The $E_7$ Gauge Enhancements And Breaking Patterns}}

\begin{eqnarray*}
	\Delta&=& z^{9} \big[-1024b_{4}^{3}+\big(((b_{1}^{2}+4b_{2})^{2}-96b_{1}b_{3})b_{4}^{2}\\&&\qquad\qquad\quad+72(b_{1}^{2}+4b_{2})b_{4}b_{6}-432b_{6}^{2}\big)z  + {\mathcal{O}}({\mathit{z}}^2) \big]
\end{eqnarray*}

\begin{table}[H]
	\begin{center}\begin{tabular}{|l|l|l|l|l|}
		\hline
		&$\deg(\Delta)$&Type&Gauge Group&Object Equation\\
		\hline
		GUT&$9$&$E_{7}$&$E_7$&$S:z=0$\\
		\hline
		Matter&$10$&$E_{8}$&$E_{8}$&$b_{4}=0$\\
		Curve&&&&\\
		\hline
	\end{tabular}
	\caption{The $E_7$ gauge enhancements. There are no
		interaction terms, and conclude that an $E_7$ GUT is not possible if we are only interested in the generic enhancements. For completeness $SU(5)$ and $SO(10)$ gauge enhancements can be found in Appendix \ref{GEs}.}
	\end{center}
	\end{table}
	
	However, we are interested in the the breaking patterns of $E_7$ which following \cite{Slansky:1981yr} are found to be
	\begin{eqnarray}\label{E7}
	(1a)\,\,E_{7} &\rightarrow& E_6\times U(1) \rightarrow  SO(10)\times U(1)^{2}\label{E71}\\
	(1b)\,\,E_{7} &\rightarrow& E_6\times U(1)\rightarrow SU(6)\times SU(2)\times U(1)\label{E72} \\
	(1c)\,\,E_{7} &\rightarrow& E_6\times U(1)\rightarrow SU(3)\times SU(3)\times SU(3)\times U(1)\\
	(2a)\,\,E_{7} &\rightarrow& SU(6) \times SU(3)\rightarrow SU(5) \times SU(3)\times U(1)\label{E73}\\
	(2b)\,\,E_{7} &\rightarrow& SU(6) \times SU(3)\rightarrow SU(4) \times SU(3) \times SU(2)\times U(1)\label{E74}\\
		(3)\,\,E_{7} &\rightarrow& SO(12)\times SU(2)\rightarrow SO(10)\times SU(2) \times U(1)\label{E75}\\
	(4)\,\,E_{7} &\rightarrow& SU(8)\label{E76}
	\label{e7break}
	\end{eqnarray}
where \ref{E71} is of key interest since under the decomposition
$$E_{7} \rightarrow E_6\times U(1)_{\delta}$$ $${\bf{133}}\rightarrow {\bf{78}}_{{0}} \oplus {\bf{1}}_{0} \oplus {\bf{27}}_{+1} \oplus \overline{{\bf{27}}}_{-1}$$
where the subscript denotes the $U(1)_\delta$ charge and under further decomposition $$E_6 \rightarrow  SO(10)\times U(1)_{\zeta}$$
\beqn
{\bf{27}} & = & {\bf{16}}_{+{1/2}} + {\bf{10}}_{-1} + {\bf{1}}_{+2},\nonumber\\
\overline{{\bf{27}}} & = & \overline{{\bf{16}}}_{-{1/2}} + \overline{\bf{10}}_{+1} + {\bf{1}}_{-2},\nonumber
\eeqn
where the subscript denotes the $U(1)_\zeta$ charge.
	
\section{The NAHE-Based Models: A Discussion}
The NAHE set consists of five basis vectors:
\begin{align*}
{\bf{1}} &= \{\psi_{\mu}^{1,2}, \chi^{1,...,6}, y^{1,...,6}, \omega^{1,...,6}|\bar{y}^{1,...,6}, \bar{\omega}^{1,...,6}, \bar{\psi}^{1,...,5}, \bar{\eta}^{1,2,3}, \bar{\phi}^{1,...,8}\}, \\
\mathbf{S} &= \{\psi_{\mu}^{1,2}, \chi^{1,...,6}\}, \\
\mathbf{b_{1}} &= \{\psi_{\mu}^{1,2}, \chi^{1,2}, y^{3,...,6}|\bar{y}^{3,...,6}, \bar{\psi}^{1,...,5}, \bar{\eta}^{1}\}, \\
\mathbf{b_{2}} &= \{\psi_{\mu}^{1,2}, \chi^{3,4}, y^{1,2}, \omega^{5,6}|\bar{y}^{1,2}, \bar{\omega}^{5,6}, \bar{\psi}^{1,...,5}, \bar{\eta}^{2}\}, \\
\mathbf{b_{3}} &= \{\psi_{\mu}^{1,2}, \chi^{5,6}, \omega^{1,...,4}  |\bar{\omega}^{1,...,4}, \bar{\psi}^{1,...,5}, \bar{\eta}^{3} \},\\
\end{align*}
The basis vectors ${\bf{1}}$ and ${\bf{S}}$, generate a model
with the $SO(44)$ gauge symmetry and ${N} = 4$
space--time supersymmetry. The vectors $\mathbf{b_i}$ for ${\mathbf{i}}=1,2,3$ correspond to the ${\mathbb{Z}}_2 \times {\mathbb{Z}}_2$ orbifold twists. The vector $\mathbf{b_{1}}$ breaks the $SO(44)$ gauge group to $SO(28)\times SO(16)$ and the ${N = 4}$
space--time supersymmetry to ${N=2}$.
The vector ${\mathbf{b_2}}$ then reduces the group to $SO(10)\times SO(22)\times SO(6)^{2}$ gauge group and the ${N = 2}$
supersymmetry is further reduced to ${N=1}$ . Furthermore, the basis vector ${\mathbf{b_3}}$ gives the decomposition $SO(10)\times SO(16)_{1}\times SO(6)^{3}$ where we fix the GGSO projection coefficient in order to preserve the ${N=1}$ space--time supersymmetry. Moreover, the sector, $\xi$, given by the linear combination
$$\xi ={ \mathbf{1+b_{1}+b_{2}+b_{3}}} \equiv \{\overline{\phi}^{1,...,8}\}$$
together with the $NS$--sector form the adjoint representation of $E_{8}$ thereby enhancing the $SO(16)_{1}$. As a result, we obtain 
$$SO(10)\times E_{8} \times SO(6)^{3}$$
as the gauge group with $N=1$ space--time supersymmetry at the NAHE level.    

There are two classes of $SO(10)$ breakings found in the literature using the free fermionic construction of the heterotic string:
\begin{itemize}
	\item Flipped $SU(5)$ \cite{Antoniadis:1988tt}, Pati-Salam \cite{Antoniadis:1990hb}, and Standard-like models \cite{Faraggi:1997dc, Faraggi:1989ka, Faraggi:1991jr, Faraggi:1991be};
	\item Left-right symmetric \cite{Cleaver:2000ds} and $SU(4)\times SU(2)\times U(1)$  models \cite{Cleaver:2002ps}.
	\end{itemize}
We are interested not in the former, but in the later case where the starting point is the $N=4$, $E_7 \times E_7 \times SO(16)$ obtained by the use of the basis $$B = \{{\bf{1}}, \bf{S}, \bf{x}, \bf{2\gamma}\}$$
where
$$\bf{x} = \{\overline{\psi}^{1,...,5}, \overline{\eta}^{1,2,3}\}$$
and
$$\bf{2\gamma}=\{\overline{\psi}^{1,2,3}, \overline{\eta}^{1,2,3}, \overline{\phi}^{1,8}\}$$
and with the following choice of one-loop GGSO projection phases

$$
\bordermatrix{~ &\bf{1}& \bf{S}& \bf{x}& \bf{2\gamma} \cr
	\bf{1} & \,\,\,\,1 & \,\,\,\,1&-1&-1 \cr
	\bf{S} & \,\,\,\,1 &\,\,\,\, 1&-1&-1\cr
	\bf{x}&-1&-1&\,\,\,\,1&\,\,\,\,1\cr
	\bf{2\gamma}&-1&-1&-1&\,\,\,\,1\cr
	}
$$
is reduced to 
$$SO(10)\times U(1)_{\zeta}\times U(1)_{\delta}\times U(1)\times SO(4)^{3}\times SO(16)$$
by introducing basis vectors  $\mathbf{b_i}$ for ${\mathbf{i}}=1,2$ corresponding to the ${\mathbb{Z}}_2 \times {\mathbb{Z}}_2$ orbifold twists.

\section{$E_{6}$ From Heterotic M-Theory}
A brief study of the existence of solutions with three generations and $E_6$ observable gauge group in the case of compactification on the torus-fibred Calabi-Yau over Hirzebruch surfaces with some remarks about the case where the base is taken to be the del Pezzo surfaces.

A nonperturbative vacuum state of the GUT theory on the observable sector is specified by a set of M-theory 5-branes wrapping a holomorphic 2-cycle on the threefold. These 5-branes are then described by a 4-form cohomology class $[W]$ satisfying the anomaly-cancellation condition being Poincar\'{e}-dual to an effective cohomology class of $H_{2}(X, \mathbb{Z})$ expressed as
$$[W]= \sigma_{\ast}(\omega)+c(F-N)+dN$$
where $c$ and $d$ are integers and $\omega$ is a class in the base manifold $B$ and $\sigma_{\ast}(\omega)$ is the pushforward to $X$ under $\sigma$ \cite{Donagi:1999ez}.

\section{The Hirzebruch Surfaces $F_{r}$}
In this section, we consider the rules for constructing realistic, viable vacua with $E_{6}$ GUT symmetry where the base manifold $B$ are taken to be the Hirzebruch surfaces, $F_{r}$. We arrive at the following conditions modified for the $E_{6}$ observable gauge group \cite{Faraggi:2002ah, Donagi:1999ez}:
\subsection{The Semistability Condition}
The semistability condition offers a choice: either 
$$\lambda\in \mathbb{Z}$$
and  $$s\,\,\text{even},\,\,e-r\,\,\text{even}$$
or 
$$\lambda = \frac{2m-1}{2},\,\,m \,\,\text{even},\quad r\,\,\text{even}.$$ 
\subsection{The Involution Conditions}
The involution conditions are 
$$\sum_{i}\kappa_{i}=\eta \cdot c_{1}(B=F_{r}) = 2e+2s-rs.$$
\subsection{The Effectiveness Condition}
The effectiveness condition boils down to
$$s \geq 24, \,\, \text{and}\,\,12r+24 \geq e$$
with 
$$\sum_{i}\kappa_{i}^{2}\leq 100+\frac{9}{4\lambda}-9\lambda$$
and 
$$\sum_{i}\kappa_{i}^{2}\leq 4+\frac{9}{4\lambda}-9\lambda+\sum_{i}\kappa_{i}.$$
\subsection{The Commutant Condition}
The commutant condition for $E_{6}$ becomes
$$\eta\geq 3c_{1}$$
which implies that
$$s\geq 6,\,\,\text{and}\,\,e\geq 3r+6.$$
\subsection{The Three Family Condition}
The three family condition reads 
$$-rs^{2}+3rs +2es -6e-6s =\frac{6}{\lambda}.$$
Solving the three family condition for $e$ assuming that the value of $s$ is known leads to
$$e(r;\lambda)  = \frac{1}{2s-6}\bigg(rs^{2}-3rs+6s+\frac{6}{\lambda}\bigg).$$
\newpage
\subsection{Space of Solutions of Type A}
In this class 
$$s\,\,\text{even},\,\,e-r\,\,\text{even},\,\,\lambda=\pm1, \pm3. $$
\begin{table}[H]
	\begin{center}
		\begin{tabular}{|c|c|}
			\hline
			$s$&$e(r; \lambda)$\\
			\hline
			&\\
			6&$3r+6+\frac{1}{\lambda}\,\,\,\,\in\mathbb{Z}$\\
			&$\lambda = \pm 1$\\
			\hline
			&\\
			8&$4r+\frac{24}{5}+\frac{3}{5\lambda}\,\,\,\,\notin\mathbb{Z}$\\
			&\\
			\hline
			&\\
			$10$&$5r+\frac{30}{7}+\frac{3}{7\lambda}\,\,\,\,\notin\mathbb{Z}$\\
			&\\
			\hline
			&\\
			12&$6r+4+\frac{1}{3\lambda}\,\,\,\,\notin\mathbb{Z}$\\
			&\\
			\hline
			&\\
			14&$7r+\frac{42}{11}+\frac{3}{11\lambda}\,\,\,\,\notin\mathbb{Z}$\\
&\\
			\hline
&\\
			16&$8r+\frac{48}{13}+\frac{3}{13\lambda}\,\,\,\,\notin\mathbb{Z}$\\
			&\\
			\hline
			&\\
			18&$9r+\frac{18}{5}+\frac{{1}}{5\lambda}\,\,\,\,\notin\mathbb{Z}$\\
			&\\
			\hline
			&\\
			20&$10r+\frac{60}{17}+\frac{3}{17\lambda}\,\,\,\,\notin\mathbb{Z}$\\
			&\\
			\hline
			&\\
			22&$11r+\frac{66}{19}+\frac{3}{19\lambda}\,\,\,\,\notin\mathbb{Z}$\\
			&\\
			\hline
			&\\
			24&$12r+\frac{24}{7}+\frac{1}{7\lambda}\,\,\,\,\notin\mathbb{Z}$\\
			&\\
			\hline
		\end{tabular}
		\caption{Solutions for $e(r;\lambda)$ for each value of $6\leq s\leq 24$ and the corresponding appropriate choice for $\lambda$ contd.}
	\end{center}
\end{table}
Clearly, the space of solutions of type A contains exactly one vacua over the Hirzebruch surfaces for any allowed value of $r$.

\subsection{Space of Solutions of Type B}
In this class 
$$r\,\,\text{even},\,\,\lambda = \pm\frac{1}{2},\pm \frac{3}{2}.$$
We will only consider the cases where integer solutions to the three family equation are found. These are given by 

$$s=6,\quad e\bigg(r;\lambda=\pm\frac{1}{2}\bigg)=3r+6+\frac{1}{\lambda}=3r+6\pm2.$$

\subsection{Explicit Expressions}
For $B=F_{r}$ 
$$\omega =(24-s)S+(12r+24-e)E$$
with
\begin{eqnarray*}
	c&=&  c_{2}+\bigg(\frac{1}{24}(n^{3}-n)+11\bigg)c_{1}^{2}-\frac{1}{2}\bigg(\lambda^{2}-\frac{1}{4}\bigg)n\eta(\eta-nc_{1})-\sum_{i}\kappa_{i}^{2}\\
	&=&100+ \frac{9}{4\lambda}-9\lambda -\sum_{i}\kappa_{i}^{2}\\
	d&=&c_{2}+\bigg(\frac{1}{24}(n^{3}-n)-1\bigg)c_{1}^{2}-\frac{1}{2}\bigg(\lambda^{2}-\frac{1}{4}\bigg)n\eta(\eta-nc_{1}) +\sum_{i}\kappa_{i}-\sum_{i}\kappa_{i}^{2} \\
	&=& 4 + \frac{9}{4\lambda}-9\lambda +\sum_{i}\kappa_{i}-\sum_{i}\kappa_{i}^{2} 	
\end{eqnarray*}
where 
$$c_{1}^{2} = 8, \quad \eta^{2}=-rs^{2}+2es,\quad \eta c_{1}=-rs+2e+2s$$
and $c_{1}$ denotes the first Chern class and $c_{2}$ denotes the second Chern class.
\section{del Pezzo Surfaces $dP_{N}$: A Word} 
We very briefly remark on the results. From \cite{Faraggi:2002ah} it can be seen that the first Chern class can be defined as
$$c_{1}(dP_{N})= \sum_{i=1}^{N}M_{i}$$
with the three family condition being
$$-2\sum_{i=1}^{N}m_{i}^{2}+n\sum_{i\neq j}m_{i}m_{j}-2n \sum_{i=1}^{N}m_{i}=\frac{6}{\lambda}$$
which for $\lambda=\pm1, \pm3$ will not admit any integer solutions and therefore no vacua exists.
\section{Conclusion}
In this paper, we looked at the $E_6$ and $E_7$ GUTs in view of the free fermionic construction of the $4D$ heterotic string.  An illustrative example, that of the NAHE-Based LRS model of \cite{Cleaver:2000ds, Cleaver:2002ps} was briefly reviewed where the starting point was taken to be the $N=4$, $E_7 \times E_7 \times SO(16)$  which was reduced to obtain the flipped $SO(10)$ GUT symmetry albeit with a different accommodation of the SM representations in the ${\bf{27}}$ of $E_{6}$. 

In light of \cite{Ashfaque:2016psv,Ashfaque:2016ydg}, we also considered the existence of solutions with three generations and $E_6$ observable gauge group. By use of Wilson lines the $E_6$ GUT symmetry can be broken to $SO(10)\times U(1)$. We found solutions of type A and solutions of type B which follow from the semistability condition, one of crucial ingredient when constructing realistic vacua contrary to \cite{Faraggi:2002ah} where flipped $SU(5)$ vacua were studied and only solutions of type B were found. It was also shown that no solutions can be found in the case of the base being the del Pezzo surfaces.

\section{Acknowledgements}
J. M. A. would like to thank String-Math 2016 organisers hosted in Paris and IGST 2016 organisers hosted in Berlin for their warm hospitality. 
\appendix

\section{{The Gauge Enhancements}}\label{GEs}
\subsection{$SU(5)$}
\begin{eqnarray*}
	\Delta &=&-z^{5}\big[P_{10}^{4}P_{5}+zP_{10}^{2}(8b_{4}P_{5}+P_{10}R)+z^{2}(16b_{3}^{2}b_{4}^{2}+P_{10}Q)\\
	&&\qquad \qquad\qquad \qquad\qquad \qquad\qquad \qquad \qquad\qquad\qquad\quad + {\mathcal{O}}({\mathit{z}}^3)\big]
\end{eqnarray*}
$$P_{10}=b_{5}$$
$$P_{5}= b_{3}^{2}b_{4}-b_{2}b_{3}b_{5}+b_{1}b_{5}^{2}$$
$$R=4b_{1}b_{4}b_{5}-b_{3}^{3}-b_{2}^{2}b_{5}$$

\begin{table}[H]
\begin{center}
	\begin{tabular}{|l|l|l|l|l|}
		\hline
		&$\deg(\Delta)$&Type&Gauge Group&Object Equation\\
		\hline
		GUT&$5$&$A_{4}$&$SU(5)$&$S:z=0$\\
		\hline
		Matter &$6$&$A_{5}$&$SU(6)$&$P_{5}: P=0$\\
		Curve&&&&\\
		\hline
		Matter&$7$&$D_{5}$&$SO(10)$&$P_{10}: b_{5}=0$\\
		Curve&&&&\\
		\hline
		Yukawa &$8$&$D_{6}$&$SO(12)$&$b_{3}=b_{5}=0$ \\
		Point&&&&\\
		\hline
		Yukawa &$8$&$E_{6}$&$E_{6}$& $b_{4}=b_{5}=0$\\
		Point&&&&\\
		\hline
		Extra &$7$&$A_{6}$&$SU(7)$& $P_{5}=R=0$,\\
		&&&&$(b_{4},b_{5})\neq (0,0)$\\
		\hline
	\end{tabular}
\end{center}
	\caption{The $SU(5)$ gauge enhancements.}
	\end{table}

\subsection{$SO(10)$}
\begin{eqnarray*}\Delta &=& -16b_{2}^{3}b_{3}^{2}z^{7}+\big(-27b_{3}^{4}-8b_{1}^{2}b_{2}^{2}b_{3}^{2}+72b_{2}b_{4}b_{3}^{2}\\&&\qquad\qquad\quad+4b_{1}b_{2}(9b_{3}^{2}+4b_{2}b_{4})b_{3}+16b_{2}^{2}(b_{4}^{2}-4b_{2}b_{6})\big)z^{8} \\&&\qquad \qquad\qquad \qquad\qquad \qquad\qquad \qquad \qquad\qquad\qquad\quad + {\mathcal{O}}({\mathit{z}}^9)\\
	&=& z^{7} \big[-16b_{2}^{3}b_{3}^{2}+\big(-27b_{3}^{4}-8b_{1}^{2}b_{2}^{2}b_{3}^{2}+72b_{2}b_{4}b_{3}^{2}\\&&\qquad\qquad\quad+4b_{1}b_{2}(9b_{3}^{2}+4b_{2}b_{4})b_{3}+16b_{2}^{2}(b_{4}^{2}-4b_{2}b_{6})\big)z \\&&\qquad \qquad\qquad \qquad\qquad \qquad\qquad \qquad \qquad\qquad\qquad\quad + {\mathcal{O}}({\mathit{z}}^2) \big]
\end{eqnarray*}

\begin{table}[H]
\begin{center}
	\begin{tabular}{|l|l|l|l|l|}
		\hline
		&$\deg(\Delta)$&Type&Gauge Group&Object Equation\\
		\hline
		GUT&$7$&$D_{5}$&$SO(10)$&$S:z=0$\\
		\hline
		Matter &$8$&$D_{6}$&$SO(12)$&$P_{10}: b_{3}=0 $\\
		Curve&&&&\\
		\hline
		Matter&$8$&$E_{6}$&$E_{6}$&$P_{16}: b_{2}=0 $\\
		Curve&&&&\\
		\hline
		Yukawa &$9$&$E_{7}$&$E_{7}$&$b_{2}=b_{3}=0$ \\
		Points&&&&$ b_{3}=b_{4}^{2}-4b_{2}b_{6}=0$\\
		\hline
	\end{tabular}
\end{center}
	\caption{The $SO(10)$ gauge enhancements.}
	\end{table}

\end{document}